\documentclass[a4paper,11pt]{article}
\pdfoutput=1 

\usepackage{jheppub} 	
\usepackage{graphicx}   	
\usepackage{dcolumn}   	
\usepackage{bm}            	
\usepackage{amssymb}   	
\usepackage{amsmath}
\usepackage{color}
\usepackage{bbold}
\usepackage{slashed}
\usepackage{epsfig}
\usepackage{epstopdf}
\usepackage[usenames,dvipsnames]{xcolor}
\usepackage{comment}
\usepackage[T1]{fontenc} 
\usepackage{color}
\begin{document}

\title{{\bf Quantum gravity effects on the thermodynamic stability of 4D Schwarzschild black hole}}

\author{Basem Kamal El-Menoufi}

\affiliation{Department of Physics and Astronomy,
University of Sussex\\
Falmer, Brighton BN1 9QH, UK}

\emailAdd{b.elmenoufi@sussex.ac.uk}

\abstract{Based on the Euclidean approach, we consider the effects of quantum gravity and mass-less matter on the thermodynamic properties of Schwarzschild black hole. The techniques of effective field theory are utilized to analytically construct the partition function at the one-loop level. Using the non-local heat kernel formalism, the partition function is expressed as a curvature expansion. We extensively discuss the effect of the corrections on the thermodynamic stability. The one-loop free energy shows, remarkably, that a large number of gauge fields is able to render Schwarzschild black hole thermodynamically stable. The black hole mass at which stability is achieved scales as $\sqrt{N}$ in Planck units, where $N$ is the number of gauge fields, and is independent of any UV completion of quantum gravity.}

\maketitle
\flushbottom

\section{Introduction}

Following the inception of black hole mechanics \cite{Bar1973}, it was soon realized their striking resemblance to the laws of thermodynamics \cite{Bek1973,Bek1974}. The subsequent discovery of Hawking radiation \cite{Haw1974} put the subject on a firm ground and made it clear that black hole thermodynamics would offer indispensable insight into quantum gravity. It is particularly important to uncover the microscopic dynamics that give rise to such macroscopic thermodynamic description. Most notably, many authors have used various approaches to try and associate a statistical origin to Bekenstein-Hawking entropy \cite{Str1996,Rov1996,Str1997}. 

One striking implication of the area law is that Schwarzschild black hole has negative specific heat. In particular, the black hole can not be in stable thermal equilibrium with radiation held at the same temperature\footnote{It is also true that all asymptotically-flat black holes can not be in thermal equilibrium \cite{Dol2014}.}. This means a statistical fluctuation would cause the hole to either grow indefinitely or evaporate away completely \cite{Haw1976}. One could nevertheless stabilize the hole {\em artificially} by placing it in an isolated finite-volume box filled with radiation whose energy is less than quarter of the hole mass \cite{Haw1976}.

In this paper we study the impact of quantum loops on the thermodynamics of 4-dimensional Schwarzschild black hole. We inquire to what extent mass-less fluctuations could affect the thermal stability of the black hole. At first sight, this might seem like a question that requires knowledge of a full theory of quantum gravity. In fact, this is not entirely true and plenty of interesting features are revealed via the semi-classical treatment of Euclidean quantum gravity. Building on earlier work\footnote{The focus in \cite{El-Men2015} was the logarithmic correction to Bekenstein-Hawking entropy which has also been discussed in a host of papers \cite{Fur1995,Gup2011,Gup20112,Sen2012,Sen2013,Car2000,Sun2001,Ban2008,Ban2009,Aro2010,Cai2010,Kau2000}.} 
\cite{El-Men2015}, we analytically construct the canonical partition function utilizing the framework of effective field theory (EFT).

Donoghue showed in \cite{Don1994} that general relativity and quantum field theory are perfectly compatible if quantum general relativity is formulated as an effective field theory (EFT). The theory is only valid below a cut-off scale, typically taken to be $M_{\text{P}}$, above which the effective description is replaced with the UV completion. The construction of the effective theory was given in \cite{Don1994,Don2012,Bur2004} and shares many practical features with chiral perturbation theory - the more familiar EFT describing the low-energy dynamics of QCD.

The main advantage of the effective theory is the ability to separate out low-energy physics. Being non-renormalizable the EFT has a poor UV behavior, yet, the unknown physics is encoded solely in the Wilson coefficients of the most general local Lagrangian. Any observable of the effective theory is expressed in terms of those {\em measured} Wilson coefficients and hence the theory is fully predictive. More interesting are the contributions induced by long-distance propagation of mass-less (light) degrees of freedom. The latter comprise reliable predictions of quantum gravity since, by the very nature of the EFT, any UV completion must reproduce these results at low energies. 

Indeed, the EFT is very well-suited to study perturbative scattering amplitudes. In a typical matrix element, long-distance physics manifests itself in non-analytic functions. At the one-loop level, the latter are finite and fully calculable in the effective theory. This insight was used in \cite{Bje2002} to determine the leading quantum correction to the Newtonian potential energy of two heavy masses. In general, low-energy predictions extracted from the EFT comprise various theorems and tests of quantum gravity \cite{Don2015}.

Clearly the gravitational effective theory must be tailored to go beyond scattering amplitudes. Various questions in quantum gravity, in particular cosmology and black hole physics, are amenable to the EFT treatment. It is then crucial to understand the structure of loop-induced modifications to classical GR in full generality. To this end, one primarily relies on the effective action as the central object and set out to consider quantum fluctuations in a fixed classical background geometry. The UV part of quantum loops induce local interactions fully encoded within the effective theory while the more interesting low-energy portion induces non-local operators in the effective action\footnote{Loop-induced non-localities and various non-local models have started to receive considerable attention with plenty of applications \cite{Esp2005,Cab2008,Woo2014,Woo2014-,Woo2007,Woo2013,Mag2016,Mag2014,Keh2014,Cal2015}.}. 

A short detour is given in section \ref{review} to review the structure of non-localities in the effective action of quantum gravity coupled to mass-less matter. As shown in \cite{El-Men2015}, the Kerr-Schild structure of Schwarzschild solution enables a precise determination of the effective action. The formalism of the non-local heat kernel expansion \cite{Bar1985,Bar1990,Avr1991} is used to express the result as an expansion in gravitational curvatures \cite{El-Men2015}. We proceed to show in detail how to compute the partition function at the one-loop level expanding around the Schwarzschild instanton. Although we only consider contributions to the partition function quadratic in curvature, the curvature expansion offers a valuable tool to decide on the structure of the partition function at higher curvature and loop-order.

The outline of the paper is as follows. After reviewing previous results in section \ref{review}, we move in section \ref{partition} to compute the free energy where we elucidate the meaning of the {\em form factor} $\ln \Box$. The free energy is used in section \ref{stability} to investigate the thermodynamic stability, primarily highlighting the impact of particle content on the fate of the black hole. Section \ref{higher} addresses the important question about the possible effects of including higher curvature and loop corrections. Section \ref{conc} is devoted to pin down some open questions that could be addressed using our formalism.

\section{Non-local effective action in Kerr-Schild spacetime: Review}\label{review}

In this section we briefly review the construction of the one-loop effective action on a fixed background Kerr-Schild (KS) geometry due to mass-less degrees of freedom. Consequently, the effective action is utilized in the next section to determine the partition function of quantum gravity at the one-loop level. Throughout we adopt the effective field theory treatment of quantum gravity \cite{Don1994,Don2012,Bur2004}. The results displayed here were derived in a previous work \cite{El-Men2015}, hence the reader familiar with the latter can skip to the next section. 

The effective action is the central object in semiclassical gravity as well as quantum gravity \cite{Bir,Par,Buc}. It elegantly encodes the effects of quantum fluctuations on the dynamics of the classical background geometry, and is computed efficiently using the background field method \cite{Abb1981}. The effective action has many applications; for example, in studying back-reaction on the classical geometry, in studying the thermal properties of black holes and in determining the UV divergences of the theory in a simple and covariant fashion.  

For a massive theory, the construction of the effective action is performed using a covariant approach primarily due to De Witt \cite{DeW1964}. In the absence of self interactions the result appears as a {\em local} expansion containing a tower of operators with ever-rising powers of derivatives of the metric, the so called DeWitt-Seeley-Gilkey expansion \cite{Avr1991,DeW1964,Gil1975}. These gravitational operators are suppressed by the mass of the matter field, in particular, all operators are analytic in derivative operators. Indeed, this expansion is not generically useful. On the one hand, the latter is completely useless when the spacetime curvature is large compared to the mass squared of the field in question. More importantly, on the other hand, mass-less fluctuations is impossible to handle using this approach.

The difficulty in dealing with operators without an intrinsic mass scale is the {\em non-locality} omnipresent in the effective action. On physical grounds, mass-less fluctuations do not have a natural IR cut-off and thus can propagate over long distances. This consequently leads to non-local behavior in the effective action: The IR portion of quantum gravity is genuinely non-local. Similar to non-analytic functions in matrix elements, these non-localities comprise reliable predictions of quantum gravity and are fully calculable using the EFT framework.  

The best available formalism to compute the effective action is based on a curvature expansion that contains {\em non-analytic} functions of derivative operators. The latter are referred to as {\em form factors} and an important example will appear below. This non-local expansion is drastically distinct from the more familiar derivative (energy) expansion described above in that it is a genuine expansion in curvatures. The main themes of the formalism were first presented in \cite{Bar1985,Bar1990}. Nevertheless, various subtleties arise in the construction that were discussed at length in \cite{El-Men2015-2} and a consistent framework was proposed.

In \cite{El-Men2015}, it was observed that working on fixed KS geometry yields an unambiguous result for the form factors. The defining property of KS spacetimes is that the full metric takes the special form \cite{Ste}
\begin{align}\label{ksmetric}
g_{\mu\nu} = \eta_{\mu\nu} - k_{\mu} k_{\nu}
\end{align}
where the vector $k^\mu = g^{\mu\nu} k_\nu$ is null with respect to both the full and flat metrics
\begin{align}
g^{\mu\nu} k_\mu k_\nu = \eta^{\mu\nu} k_\mu k_\nu  =0 \ \ .
\end{align}

It is important to stress here that the above comprises a covariant ansatz for the metric, i.e. it does not depend on the existence of a special coordinate system. Nevertheless, recognizing a spacetime to be of the KS type is indeed easier in certain coordinates than others. It is quite remarkable that both Schwarzschild and Kerr solutions fall in this class. Thanks to the KS ansatz, Roy Kerr was able to obtain his celebrated solution in closed form \cite{Ker1963,Sch,Deb1969}. Here, we are primarily concerned with Schwarzschild black hole where the null vector, in standard Cartesian coordinates, takes the form
\begin{align}\label{KSvector}
k_\mu = \sqrt{\frac{2GM}{r}} \left(1,\frac{\bold{x}}{r}\right) \ \ .
\end{align}

When substituted back into eq. (\ref{ksmetric}), the reader might recognize the Schwarzschild metric as expressed by Eddington. A detailed solution of Einstein equation starting with KS ansatz was first given in \cite{Adl1973} and reviewed in \cite{El-Men2015}.

After integrating out the matter sector and graviton at one-loop, the effective action is composed of two pieces
\begin{align}
\Gamma[\bar{g}] = \Gamma_{\text{local}}[\bar{g};\mu] + \Gamma_{\text{ln}}[\bar{g};\mu]
\end{align} 
where $\bar{g}$ refers to the background KS metric. The local contribution is nothing but the {\em renormalized} EFT action\footnote{It is of particular importance to realize that Newton's constant does not get renormalized when one considers only mass-less fields. Consistent with the EFT power-counting, the one-loop divergences are proportional to $R^2$ terms.}
\begin{align}\label{localaction}
\nonumber
\Gamma_{\text{local}}[\bar{g};\mu] = \int d^4x \, &\bigg( \frac{M_{\text{P}}^2}{2} \, R + c^r_1(\mu) \, R^2 +c^r_2(\mu) \, R_{\mu\nu} R^{\mu\nu}\\
&+ c^r_3(\mu) \, R_{\mu\nu\alpha\beta} R^{\mu\nu\alpha\beta} + c^r_4(\mu)\, \nabla^2 R +\mathcal{O}\left(R^3\right) \bigg)
\end{align}
where the renormalized couplings are scale-dependent. The other piece is non-local representing the long-distance portion of quantum loops
\begin{align}\label{nonlocalaction}
\nonumber
\Gamma_{\text{ln}}[\bar{g};\mu] = - \int d^4x \, &\bigg(\alpha\, R \ln\left(\frac{\Box}{\mu^2}\right) R + \beta\, R_{\mu\nu} \ln\left(\frac{\Box}{\mu^2}\right) R^{\mu\nu} \\
&+ \gamma\, R_{\mu\nu\alpha\beta} \ln\left(\frac{\Box}{\mu^2}\right) R^{\mu\nu\alpha\beta} + \Theta\, \ln\left(\frac{\Box}{\mu^2}\right) \Box R + \mathcal{O}\left(R^3\right) \bigg)
\end{align}
where $\Box$ is the flat-space d' Alembertian and the operator $\ln\left(\Box\right)$ is our first example of a form factor. Notice here that we only consider the action up to second order in curvature. Section \ref{higher} is devoted to discuss the effects of including higher curvature corrections. Listed in table \ref{tablecoeff}, the different coefficients are spin-dependent and come straight out of the computation. 
\begin{table}
\centering
\begin{tabular}{c|c|c|c|c}
\hline
\hline
  & $\alpha$ & $\beta$ & $\gamma$ & $\Theta$ \\
\hline
\text{Scalar} & 5 & -2 & 2 & -6 \\  
\text{Fermion} & -5 & 8 & 7 & -- \\
U(1)\text{boson} & -50 & 176 & -26 & -- \\
\text{Graviton} & 430 & -1444 & 424 & -- \\
\hline
\hline 
\end{tabular}
\caption{The coefficients appearing in the effective action due to massless fields of various spins \cite{El-Men2015}. All numbers are divided by $11520 \pi^2$.}
\label{tablecoeff}
\end{table}

The renormalization group (RG) properties of the effective action will be of most importance later. Indeed, the renormalized action is invariant under RG flow by virtue of the beta function of the various coefficients \cite{El-Men2015}
\begin{align}\label{RG}
\nonumber
c^r_1(\mu) &= c^r_1(\mu_\star) - \alpha \ln \left(\frac{\mu^2}{\mu_\star^2}\right) & c^r_2(\mu) &= c^r_2(\mu_\star) - \beta \ln \left(\frac{\mu^2}{\mu_\star^2}\right)\\
c^r_3(\mu) &= c^r_3(\mu_\star) - \gamma \ln \left(\frac{\mu^2}{\mu_\star^2}\right) & c^r_4(\mu) &= c^r_4(\mu_\star) - \Theta \ln \left(\frac{\mu^2}{\mu_\star^2}\right)  \ \ .
\end{align}

\section{Free energy of Schwarzschild black hole}\label{partition}

In this section, we utilize the results from last section to compute the partition function for the canonical ensemble of quantum gravity coupled to mass-less matter. The construction commences analogously to thermal field theory \cite{Gib1994,Gib1976,Gro1982}
\begin{align}\label{Zdef}
Z(\beta) = \int \mathcal{D} \Psi \mathcal{D} g \, e^{-\mathcal{S}_E - \mathcal{S}_{\partial}} \ \ .
\end{align}
Here, $\Psi$ is any matter field, $g$ is a Euclidean metric, $\mathcal{S}_E$ is the Euclidean action of the theory and $\mathcal{S}_{\partial}$ is the Hawking-Gibbons-York boundary action given by \cite{Gro1982,Yor1972}
\begin{align}
\mathcal{S}_{\partial} = -\frac{1}{8\pi G} \int_{\partial \mathcal{M}}\, \sqrt{\gamma} (K - K_0) \ \ .
\end{align}
The boundary action ensures the variational problem is well posed. It clearly does not contribute to the equations of motion but nevertheless plays a crucial role in black hole thermodynamics. To define the canonical ensemble, one has to impose boundary conditions on the variables of the path integral. The prescription is to integrate over positive-definite metrics\footnote{The asymptotic behavior of such metrics is determined by requiring finite action \cite{Gro1982}.} which approach the flat metric on ${\rm I\!R}^3 \times S^1$ \cite{Gib1994,Gib1976,Gro1982}. The matter fields appearing in the path integral are subject to boundary conditions appropriate for the canonical ensemble.

Among the various attempts to discuss the thermal properties of black holes, the Euclidean approach stands out as a self-consistent framework. In a semi-classical evaluation of the partition function, the black hole appears as an extremal point of the Euclidean action yielding a non-trivial contribution to the partition function. Most notably, the Euclidean approach was successful in reproducing the Bekenstein-Hawking entropy for Schwarzschild black hole \cite{Gib1976}. Moreover, the formalism is readily extended to handle rotating and charged black holes by constructing the grand canonical partition function as well as (A)dS black holes \cite{Haw1982}.

We evaluate the partition function using the semi-classical ($\hbar \rightarrow 0$) approximation by expanding around a background metric which extremizes the action
\begin{align}\label{metricdecomp}
g_{\mu\nu} = \bar{g}_{\mu\nu} + \kappa\, h_{\mu\nu}
\end{align}
where $ \bar{g}_{\mu\nu}$ is also known as a gravitational instanton, $h_{\mu\nu}$ is the metric fluctuation and $\kappa = \sqrt{32 \pi G}$. If $\bar{g}$ is a KS space-time, then one could readily perform an analytic continuation on the effective action to obtain the partition function at the one-loop level. This yields \cite{El-Men2015}
\begin{align}\label{partitionfunc}
\ln Z = \Gamma_{\text{local}}[\bar{g}] + \Gamma_{\text{ln}}[\bar{g}]  - \mathcal{S}_{\partial}
\end{align} 
where now
\begin{align}\label{logspecies}
\nonumber
\Gamma_{\text{ln}}[\bar{g}] = - \int d^4x \, &\bigg[\alpha R \ln\left(\frac{-\Delta}{\mu^2}\right) R + \beta R_{\mu\nu} \ln\left(\frac{-\Delta}{\mu^2}\right) R^{\mu\nu} \\
&+ \gamma R_{\mu\nu\alpha\beta} \ln\left(\frac{-\Delta}{\mu^2}\right) R^{\mu\nu\alpha\beta} - \Theta \ln\left(\frac{-\Delta}{\mu^2}\right) \Delta R \bigg] \ \ ,
\end{align}
and $\Delta$ being the $4\text{D}$ flat Laplacian on ${\rm I\!R}^3 \times S^1$. We truncated the partition function at second order in the curvature expansion and devote section \ref{higher} to discuss possible effects of higher curvature contributions.

	\subsection{Comments on the gravitational instanton}
	
	Here we provide important comments regarding the gravitational instanton appropriate for our case. Let us start by using eq. (\ref{KSvector}) to write the full Schwarzschild metric
	\begin{align}
	ds^2 = \left(1-\frac{2GM}{r}\right) dt^2_{\star} - \left(1+\frac{2GM}{r}\right) dr^2 - \frac{4GM}{r} dt_\star dr - r^2 d\Omega_2
	\end{align}
	where $t_\star$ is the KS time coordinate related to the usual Schwarzschild time as $t = t_\star  + 2GM \ln (2GM/r -1)$. There are two important features: the metric is both regular across the horizon and not static when expressed in standard spherical (Cartesian) coordinates of the flat metric.
	
	According to our formalism, the background instanton in eq. (\ref{logspecies}) is defined as usual via the analytic continuation $t_\star \to -i \tau$, where the Euclidean time is periodic defining the temperature of the canonical ensemble. Doing so one sees immediately that the resulting metric is complex, i.e. quasi-Euclidean. This is different from the familiar real Euclidean section obtained by continuing the Schwarzschild time coordinate \cite{Gib1976}. Nevertheless, this situation is customary for any stationary spacetime and pauses no concern at all. Within the Euclidean approach, Brown et. al. have elucidated the use of complex geometries in deriving black hole thermodynamics\footnote{In this regard, see also the discussion in \cite{Mon2010}.} \cite{Bro1991}. As we shall see, the instanton action is real and recovers all thermodynamic relations derived using the real Euclidean section. 
				
	It is remarkable that the resulting metric, albeit complex, is {\em positive-definite}. An immediate consequence is that $r=2GM$ is a fixed point of the Killing field $\partial/\partial \tau$ and the space ends there as usual. On the other hand, the regularity across the horizon implies that the period of Euclidean time can {\em not} be fixed because the metric does not possess a conical singularity. Nevertheless, it is fully consistent to fix the temperature by simply envoking the zeroth law of black hole thermodynamics, i.e. $T = \kappa/2\pi$, where $\kappa$ being the surface gravity of the hole.

	\subsection{What exactly is the logarithm?}
	
It is clear that to evaluate the partition function one needs to understand how the operator $\ln \Delta$ is represented in position space. This is our task in the current section. First of all, one needs to impose the appropriate thermal boundary conditions on matter and gravitational fluctuations. A useful way to accomplish this is by implementing the following identity for the logarithm
\begin{align}
\ln\left(\frac{-\Delta}{\mu^2}\right) = - \int_0^\infty \, dm^2 \left[(-\Delta + m^2)^{-1} - (\mu^2 + m^2)^{-1} \right] \ \ .
\end{align}
For a boson (fermion) field, the inverse operator is interpreted to be the 2-point function subject to periodic (anti-periodic) boundary conditions. For example, we have
\begin{align}\label{thermal2pt}
G_{\text{bosons}} (\tau,\vec{x}) = \beta^{-1} \sum_n \int \frac{d^3q}{(2\pi)^3} \frac{e^{i(\omega_n \tau + i \vec{q} \cdot \vec{x})}}{\omega_n^2+q^2+m^2}, \quad \quad \omega_n = \frac{2\pi n}{\beta} \ \ .
\end{align}
To speak of thermodynamic equilibrium, one only considers {\em stationary} metrics as saddle points. In this case, integration over the thermal circle is readily done at the onset. Using eq. (\ref{thermal2pt}) in eq. (\ref{logspecies}), one gets
\begin{align}
\nonumber
\Gamma_{\text{ln}}[\bar{g}] = - \beta \int d^3x d^3x^\prime \, &\bigg[\alpha R \mathcal{L}(\vec{x}-\vec{x}^\prime) R + \beta R_{\mu\nu} \mathcal{L}(\vec{x}-\vec{x}^\prime) R^{\mu\nu} \\
&+ \gamma R_{\mu\nu\alpha\beta} \mathcal{L}(\vec{x}-\vec{x}^\prime) R^{\mu\nu\alpha\beta} - \Theta \mathcal{L}(\vec{x}-\vec{x}^\prime) \Delta R \bigg] \ \ .
\end{align}
where
\begin{align}
\mathcal{L}(\vec{x}-\vec{x}^\prime) \equiv \ln \left(\frac{-\nabla^2}{\mu^2}\right) \delta^{(3)}(\vec{x}-\vec{x}^\prime)
\end{align}
The character of the operator is better revealed in momentum space. Setting $\vec{x}^\prime =0$ for simplicity, we get
\begin{align}
\mathfrak{L}(\vec{x}) = \int \frac{d^3q}{(2\pi)^3} \ln \left(\frac{\vec{q}^2}{\mu^2}\right) e^{i \vec{q} \cdot \vec{x}} \ \ .
\end{align}
As it stands, the above integral is ill-defined due to the rapid oscillation of the exponential factor at $|\vec{q}| \rightarrow \infty$. Even more problematic is the apparent short-distance divergence of the integral, i.e. when $\vec{x} \rightarrow 0$. We shall now present an appropriate method to define the non-local kernel - $\mathfrak{L}(\vec{x})$ -  as a distribution. One starts by imposing a regulator as follows
\begin{align}\label{logactingdelta}
\mathfrak{L}(\vec{x}) = \lim_{\epsilon \rightarrow 0} \int \frac{d^3q}{(2\pi)^3} \ln \left(\frac{\vec{q}^{2}}{\mu^2}\right) e^{i \vec{q} \cdot \vec{x}} e^{-\epsilon q} \ \ .
\end{align}
The sign of the exponent in the regulator is dictated by Feynman's prescription, i.e.  $\vec{x}^2 \to \vec{x}^2 + i\epsilon$. The angular integrals are readily done and three structures emerge
\begin{align}
\mathfrak{L}(\vec{x}) = \left( \mathfrak{L}_1 (\vec{x}) + \mathfrak{L}_2 (\vec{x}) - \mathfrak{L}_3 (\vec{x}) \right) - \ln \mu^2\, \delta^{(3)}(\vec{x})
\end{align}
where
\begin{align}
\nonumber
\mathfrak{L}_1 (\vec{x}) &= \frac{i(\gamma_E-1)}{2\pi^2 r^2} \lim_{\epsilon \rightarrow 0} \left(\frac{1}{r-i\epsilon} - \frac{1}{r + i\epsilon} \right) \\\nonumber
\mathfrak{L}_2 (\vec{x}) &= \frac{i \ln r}{2\pi^2 r^2} \lim_{\epsilon \rightarrow 0} \left(\frac{1}{r-i\epsilon} - \frac{1}{r+i\epsilon}\right)\\
\mathfrak{L}_3 (\vec{x}) &=  \frac{1}{4\pi r^2} \lim_{\epsilon \rightarrow 0}  \left(\frac{1}{r-i\epsilon} + \frac{1}{r+i\epsilon}\right) \ \ .
\end{align}
Each term in the above is a well-defined distribution as we now show. Notice that both $\mathfrak{L}_1 (\vec{x})$ and $\mathfrak{L}_2 (\vec{x})$ vanish identically unless $r=0$. Integrating against a smooth function one finds
\begin{align}
\mathfrak{L}_1 (\vec{x}) = 2 (1-\gamma_E) \delta^{(3)}(\vec{x}), \quad\quad \mathfrak{L}_2 (\vec{x}) = -2 \lim_{\epsilon \rightarrow 0} (\ln \epsilon) \, \delta^{(3)}(\vec{x}) \ \ .
\end{align}
The last piece - $\mathfrak{L}_3 (\vec{x})$ - is straightforward as it would yield the principal-value of an integral. Finally, one obtains\footnote{See \cite{Xav2017} for an alternative derivation.}
\begin{align}\label{logfinal}
\mathfrak{L}(\vec{x}-\vec{x}^\prime) = -\frac{1}{2\pi} \lim_{\epsilon \rightarrow 0} \left[\mathcal{P}\left(\frac{1}{|\vec{x}-\vec{x}^\prime|^3}\right) + 4\pi (\ln(\mu\epsilon)+\gamma_E-1) \delta^{(3)}(\vec{x}-\vec{x}^\prime)\right] \ \ .
\end{align}
The precise application of the principal-value and the limiting procedure will be crucial to obtain a well-behaved finite result when the non-local function is integrated over in the partition function.

	\subsection{Exact evaluation of the partition function}

Now that we are armed with an exact expression for the logarithm, we proceed to evaluate the partition function of Schwarzschild black hole. In this case, only the piece involving the Riemann tensor in eq. (\ref{logspecies}) contributes. It will suffice to present an explicit computation using only few components of the Riemann tensor, which are determined easily for Schwarzschild metric using eq. (\ref{KSvector}). For instance, consider the following structure\footnote{The super(sub)-script refers to the power of $G$.}
\begin{align}
\nonumber
\overset{\scriptscriptstyle{(2)}}{\Gamma}_{\text{ln}} \subset \int d^3x\, d^3x^\prime \overset{\scriptscriptstyle{(1)}}{R}_{ittj}(x) \, \mathfrak{L}(\vec{x}-\vec{x}^\prime) \,\underset{\scriptscriptstyle{(1)}}{R}^{ittj}(x^{\prime}) 
\end{align}
where both indices $(ij)$ are being summed over and
\begin{align}
\overset{\scriptscriptstyle{(1)}}{R}_{ittj}(x) = \frac{GM}{r^3} \left(\frac{3x_i x_j}{r^2} - \delta_{ij}\right) \ \ .
\end{align}
The non-local kernel, eq. (\ref{logfinal}), give rise to two integrals as follows
\begin{align}
\text{I}(M,\epsilon) &= -\frac{1}{2\pi} \int d^3x\, d^3x^\prime \overset{\scriptscriptstyle{(1)}}{R}_{ittj}(x) \mathcal{P}\left(\frac{1}{|\vec{x}-\vec{x}^\prime|^3}\right)\underset{\scriptscriptstyle{(1)}}{R}^{ittj}(x^{\prime}) \\
\text{J}(\mu,\epsilon) &= -2 (\ln(\mu\epsilon)+\gamma_E-1) \int d^3x \, \overset{\scriptscriptstyle{(1)}}{R}_{ittj} \underset{\scriptscriptstyle{(1)}}{R}^{ittj} \ \ .
\end{align}
The second equation above is a local contribution readily evaluated
\begin{align}
\text{J}(\mu,\epsilon) = -\frac{2\pi}{GM}\, \lim_{\epsilon \to 0} (\ln(\mu\epsilon) + \gamma_E -1) \ \ .
\end{align}
Moving on to the first integral we find
\begin{align}\label{theIintegral}
\text{I}(M,\epsilon) = -\frac{3(GM)^2}{2\pi} \int d^3x\, d^3x^\prime\,  \mathcal{P}\left(\frac{1}{|\vec{x}-\vec{x}^\prime|^3}\right) \frac{3(\vec{x} \cdot \vec{x}^\prime)^2 - (r r^\prime)^2}{(r r^\prime)^2} \  \ .
\end{align}
With the above form, it is advisable to switch to spherical polar coordinates. Here comes the most important point of the derivation: how to obtain the principal-value of the above integral? We devise the prescription of imposing the principal-value on the radial portion of the integral in eq. (\ref{theIintegral}) and show that it yields a well-behaved result. Without loss of generality, if we first evaluate the $d^3x^\prime$ integral then spherical symmetry allows us to align the $z^\prime$-axis along $\vec{x}$. One finds
\begin{align}
\nonumber
\text{I}(M,\epsilon) = - 3 (GM)^2\, \lim_{\epsilon \to 0} \int \frac{d^3x}{r} \bigg[&\int_{2GM}^{r-\epsilon} \frac{dr^\prime}{r^\prime} \int_{-1}^{1}du \frac{3u^2-1}{(r^2 + r^{\prime 2} - 2r r^\prime u)^{3/2}} \\
+&\int_{r+\epsilon}^{\infty} \frac{dr^\prime}{r^\prime} \int_{-1}^{1}du \frac{3u^2-1}{(r^2 + r^{\prime 2} - 2r r^\prime u)^{3/2}} \bigg] \ \ .
\end{align}
The above step explicitly describes our prescription. Performing the $du$ integral, the remaining integrals are fully regularized and one finds
\begin{align}
\text{I}(M,\epsilon) = - \frac{2\pi}{GM} \left[\ln\left(\frac{2GM}{\epsilon}\right) + \text{constant}\right] \ \ .
\end{align} 
Putting everything together, the factor $\ln \epsilon$ cancels out leaving a finite result as promised
\begin{align}\label{mulesson}
\overset{\scriptscriptstyle{(2)}}{\Gamma}_{\text{ln}} \subset -\frac{2\pi}{GM} \, \ln (2GM\mu) \ \ .
\end{align}
Here, we absorbed all numerical constants in $\ln \mu$ which amounts to a finite renormalization. Of most importance is the following lesson: $\Gamma_{\text{ln}}$ must be proportional to $\ln (2GM\mu)$. It is then straightforward to finish the computation in the manner we just described\footnote{In fact, looking at eq. (\ref{mulesson}) one can extract the final result by simply computing the coefficient of $\ln \mu$ in the partition function.}. Taking into account the local contributions from the bulk and boundary terms, one reaches the partition function at the one-loop order
\begin{align}
\ln Z(\beta) = - \frac{\beta^2}{16\pi G} + 64\pi^2 \bigg[c_3(\mu) + 2 \, \Xi \ln(\mu \beta)\bigg]
\end{align} 
where we used $\beta = 8\pi G M$ and
\begin{align}\label{sumparticles}
\Xi = \frac{1}{11520\pi^2} \left(2 N_s + 7 N_f - 26 N_V + 424 \right) 
\end{align}
counts the contribution of various species. Using this expression one can immediately recover the logarithmic correction to Bekenstein-Hawking entropy \cite{El-Men2015,Fur1995,Gup2011,Gup20112,Sen2012,Sen2013,Car2000,Sun2001,Ban2008,Ban2009,Aro2010,Cai2010,Kau2000}. Notice here that our quasi-Euclidean instanton yields the correct tree-level partition function.
	
	\subsection{Free energy: scale independance and dimensional transmutation}
	
With the partition function at our disposal, one proceeds to write down the free energy of Schwarzschild black hole
\begin{align}\label{fenergyinitial}
F(\beta) = \frac{\beta}{16 \pi G} - \frac{64 \pi^2 }{\beta}\bigg(c_3(\mu) + 2\, \Xi \ln(\mu \beta) \bigg) \ \ .
\end{align}
Notice in particular that the free energy is invariant under RG flow which must be the case since it is a physical quantity. A simple computation shows
\begin{align}
\frac{d}{d\ln \mu} F(\beta) = 0
\end{align}
using the RG equation of the Wilson coefficient eq. (\ref{RG}). Hence, dimensional transmutation allows us to trade off the constant $c_3$ with a mass scale via
\begin{align}
c_3(\mu)  = -2\, \Xi \ln (\mu \beta_{\text{QG}}) \ \ .
\end{align}
Substituting back in eq. (\ref{fenergyinitial}), we arrive at
\begin{align}\label{fenergyfinal}
F(\beta) =  \frac{\beta}{16 \pi G} - \frac{128 \pi^2 \Xi }{\beta} \ln\left(\frac{\beta}{\beta_{\text{QG}}}\right) 
\end{align} 	
which is manifestly free of the unphysical scale $\mu$. The scale $\beta_{\text{QG}}$ is intrinsically tied to the UV completion of quantum gravity. The constant $c_3$,  and thus $\beta_{\text{QG}}$, could in principle be determined either by matching onto the full theory at some scale or using experimental input. Armed with an expression for the free energy, the next section is devoted to investigate the thermodynamic stability.

\section{Thermodynamic stability}\label{stability}

In classical thermodynamics, an isolated system would be in stable equilibrium if and only if the entropy is a concave function of all extensive parameters \cite{Cal1985}. This follows simply from the second law of thermodynamics and basic properties of the entropy. Consider a random fluctuation that transfers an amount of energy $\Delta U$ from one part of the system, say initially containing energy $\lambda U$, to the rest of the system. If this process increases the entropy, then it is allowed by the second law. Suppressing dependance on further extensive variables, stability then requires
\begin{align}   
S(\lambda U - \Delta U) + S((1-\lambda) U + \Delta U) \leq S(U)
\end{align}
where $0 \leq \lambda \leq 1$. The entropy is assumed to be a first-order homogeneous function, and thus one easily uncovers the condition of concavity\footnote{We defined $U_1 = U-\Delta U/\lambda$ and $U_2 = U-\Delta U/(1-\lambda)$.} 
\begin{align}
\lambda S(U_1) + (1-\lambda) S(U_2) \leq S(\lambda U_1 + (1-\lambda) U_2) \ \ .
\end{align}
This condition comprises a {\em global} criterion of stability since the size of the fluctuation $\Delta U$ is arbitrary. Reinstating all extensive variables one finds analogously that stability requires the entropy to be a concave function of all extensive variables. This requires the Hessian of the entropy to be positive definite, and in particular
\begin{align}
\frac{\partial^2 S}{\partial U^2} \leq 0
\end{align}
which implies the positivity of the heat capacity in stable systems. 

One can now inquire about the stability of a system exchanging energy with a heat reservoir at constant temperature. A series of steps shows that the Helmholtz free energy must satisfy \cite{Cal1985,Pre2003}
\begin{align}\label{stabcondition}
\frac{\partial^2 F}{\partial T^2} < 0 
\end{align}
which derives directly from the fact that the system plus the reservoir comprise an isolated system whose stability requires the total entropy to be concave.

A close look at eq. (\ref{fenergyfinal}) reveals that the sign of $\Xi$ plays the dominant role in determining the stability of the black hole. This is controlled by the particle content of the theory.    

\subsection{Case I: $\Xi < 0$}

\begin{figure}
\center
  \includegraphics[scale=0.35]{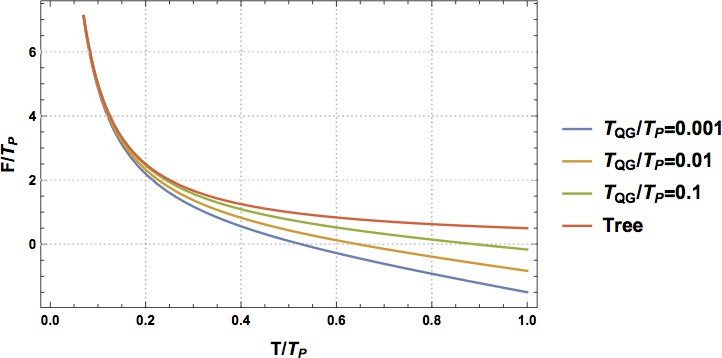}
  \caption{The free energy as a function of the temperature of Schwarzschild black hole at tree level and including the one-loop contribution. Without loss of generality, we exclude the graviton and set $N_s=N_f=0$.}
  \label{fscale2}
\end{figure}

\begin{figure}
\center
  \includegraphics[scale=0.35]{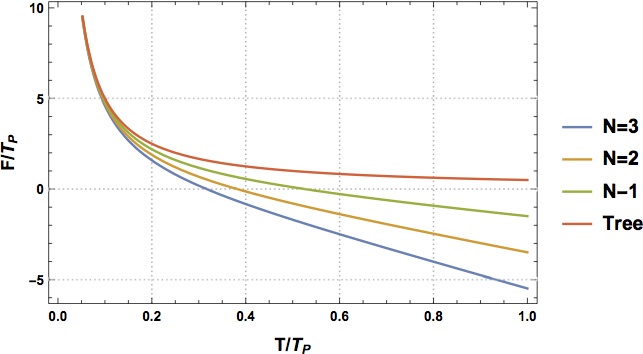}
  \caption{The free energy as a function of the temperature of Schwarzschild black hole at tree level and including the one-loop contribution. Without loss of generality, we exclude the graviton and set $N_s=N_f=0$. Here, we fix $T_{\text{QG}}/T_{\text{P}} = 0.001$.}
  \label{fnumber2}
\end{figure}

This case emerges in a theory with large number of gauge fields. For example, if we neglect the graviton contribution then the minimal standard model falls in this class. Rescaling the free energy by the Planck temperature $T_{\text{P}}  = (8\pi G)^{-1/2}$, we show various plots of the free energy in figures \ref{fscale2} and \ref{fnumber2}. In particular, we observe that lowering the scale $T_{\text{QG}}$ has the same effect as increasing the number of gauge fields. Let us define $\Sigma = | \Xi |$ and compute the second derivative of the free energy
\begin{align}
\frac{\partial^2 F}{\partial T^2} = \frac{(8\pi G)^{-1}}{T^3} - \frac{128 \pi^2 \Sigma}{T} \ \ .
\end{align}
Clearly, this becomes negative at a critical temperature
\begin{align}\label{Tc}
T_{\text{C}} = \sqrt{\frac{90}{(26 N_V - 7 N_f - 2N_s - 424)}} \, T_\text{P}  \ \ .
\end{align}
The quantum correction, accurate up to quadratic order in curvatures, is able to render the black hole thermodynamically stable. In the large-$N_{V}$ limit, the critical temperature could be made parametrically smaller than the Planck temperature. One remarkable feature of this result is that it is insensitive to the UV scale $T_{\text{QG}}$. Only the low-energy content of the theory is needed to decide on the possible thermodynamic stability of Schwarzschild black hole.


\subsection{Case II: $\Xi > 0$}

\begin{figure}
\center
  \includegraphics[scale=0.35]{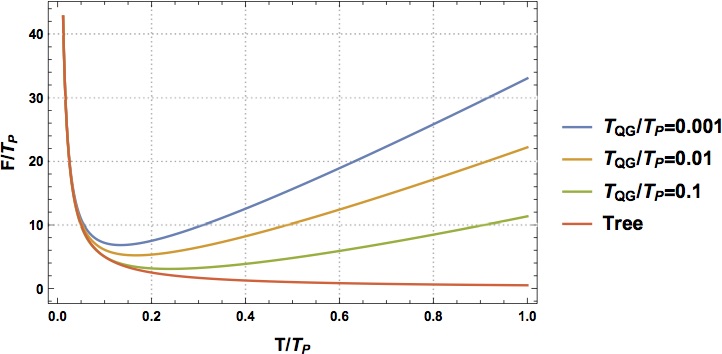}
  \caption{The free energy as a function of the temperature of Schwarzschild black hole at tree level and including the one-loop contribution. Without loss of generality, only the contribution of the graviton to $\Xi$ is considered.}
  \label{fscale}
\end{figure}

\begin{figure}
\center
  \includegraphics[scale=0.35]{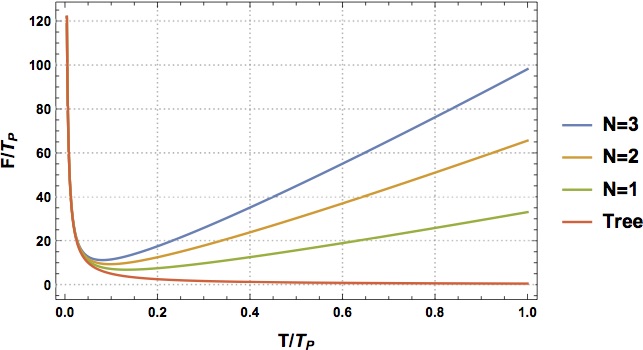}
  \caption{The free energy as a function of the temperature of Schwarzschild black hole at tree level and including the one-loop contribution. Without loss of generality, only the contribution of the graviton to $\Xi$ is considered. Here, N is an overall multiplicative factor to mimic the effect of increasing $N_s$ ($N_f$) and $T_{\text{QG}}/T_{\text{P}} = 0.001$.}
  \label{fnumber}
\end{figure}

This case is particularly interesting as it arises if the particle content is that of the standard model plus a single graviton. Using eq. (\ref{stabcondition}) we clearly see that thermal stability is not attained in this case. Various plots of the free energy are shown in figures \ref{fscale} and \ref{fnumber}, where we scaled the free energy by the Planck temperature $T_{\text{P}}  = (8\pi G)^{-1/2}$. The main observation is that the loop correction forces the free energy to develop a minimum. Both the position of the minimum and the value of the free energy are controlled by the number of particles as well as the scale $T_{\text{QG}}$. 

The existence of a minimum in the free energy has important consequences because it corresponds to the vanishing of the entropy. The entropy reads
\begin{align}
\nonumber
S_{\text{bh}}(T) &= -\frac{\partial F}{\partial T}  \\
&= \frac{T^2_\text{P}}{2T^2} + 128 \pi^2 \Xi \left( \ln \frac{T_{\text{QG}}}{T} -1 \right) \ \ ,
\end{align}
which vanishes at the solution to the following transcendental equation
\begin{align}
\frac{T}{T_{\text{QG}}} = e^{\left(\lambda/T^2-1\right)}, \quad \lambda = \frac{T^2_{\text{P}}}{256 \pi^2 \Xi} \ \ .
\end{align}
Recalling that $T_{\text{QG}} > 0$, the above equation must admit one solution. The crucial aspect of this result is manifest in figure \ref{fscale}. If the scale $T_{\text{QG}}$ is low enough in Planck units, the temperature at which the entropy vanishes is parametrically sub-Planckian. We offer few remarks about this result in section \ref{conc}.


\section{Higher curvature and loops}\label{higher}

All results obtained so far were derived using the partition function accurate up to second order in curvatures. It is crucial to understand {\em if} and {\em how} higher curvature corrections could change the main conclusions drawn from the analysis. Here we have to distinguish between the gravitational and matter sectors. One one hand, the latter is one-loop exact and one can accurately determine the effect of the missing corrections. On the other, matter contribution is dominant in the large-N limit and one can safely ignore quantum gravity.

	\subsection{Matter sector}
	Higher order corrections in this case are purely non-local. Only the $\mathcal{O}(R^2)$ terms of the renormalized EFT action, eq. (\ref{localaction}), receive contribution from mass-less minimally coupled (MMC) fields. One could understand this on dimensional grounds: divergences from MMC fields can only be proportional to quadratic curvature invariants due to the absence of a mass scale in the problem. However, higher curvature non-local operators in eq. (\ref{nonlocalaction}) indeed receive contributions from the matter sector. An example of an operator possibly appearing at cubic order reads
	
	\begin{align}\label{nonlocalcubic}
	\ln Z \subset \int d^4x \, R_{\mu\nu}^{~~~\rho\sigma} R_{\rho\sigma}^{~~~\gamma\beta} \frac{1}{\Box} R_{\gamma\beta}^{~~~\mu\nu} \ \ .
	\end{align}
	One can determine the effect of the above operator by observing the scaling properties of the partition function. In particular, eq. (\ref{nonlocalcubic}) is invariant under a scale transformation of the background metric $\bar{g} \to \Lambda^2 \bar{g}$ and hence contributes at most a numerical constant to $\ln Z$. Hence, such an operator has no effect on the stability criterion, eq. (\ref{stabcondition}), albeit changing the entropy by an inconsequential constant. Extrapolating this analysis, one can deduce that all higher curvature non-local operators are likewise scale invariant and thus bear no consequence on the stability of the black hole. 
	
	\subsection{Gravitational sector}
		The story is more complicated in this case. Although the previous analysis still holds true for pure gravity at the one-loop level, the structure of the partition function is richer at two loops and beyond. For example, according to the EFT power counting, the divergences of two-loop quantum gravity are proportional to $\mathcal{O}(R^3)$ and have been determined long ago in \cite{Gor1985}. Hence, a new local operator in eq. (\ref{localaction}) would be renormalized\footnote{Here, we express the operator for KS backgrounds ($\sqrt{g} =1$).} 
		\begin{align}
		\ln Z \subset \frac{d^r(\mu)}{M_P^2} \int d^4x \, R_{\mu\nu}^{~~~\rho\sigma} R_{\rho\sigma}^{~~~\gamma\beta}  R_{\gamma\beta}^{~~~\mu\nu} \ \ .
		\end{align}
		Once again, we can use scaling arguments to understand the effect induced by such an operator. Under the transformation $\bar{g} \to \Lambda^2 \bar{g}$, one simply sees that $\ln Z$ picks a non-trivial correction proportional to $T^2/T_{\text{P}}^2$. As expected, the effect of higher curvature {\em local} operators is negligible below the Planck temperature.
		
		Next, how about the non-local operators generated from higher loops? Without further computation, it is difficult to determine the exact form of these corrections. Nevertheless, RG invariance enables us to make a strong statement: the logarithmic correction in eq. (\ref{fenergyinitial}) is exact to any loop order. Following the effective theory power counting, the beta function of the Wilson coefficients, eq. (\ref{RG}), does not receive any contributions from higher graviton loops. Hence, the analysis we presented is quite robust especially if one invokes a large number of matter fields. 		
				
\section{Discussion}\label{conc}

A full understanding of black hole thermodynamics could be our guide to uncover a consistent theory of quantum gravity. Even though this goal seems to be far ahead, various questions could be addressed employing effective field theory techniques. Quantum GR is a self-consistent effective theory valid up to the Planck scale. Most importantly, the effective theory enables us to extract reliable predictions from the low-energy portion of quantum loops. The latter manifests in non-local operators in the effective action, or likewise the partition function, while the unknown UV physics is encoded in the renormalized Wilson coefficients. 

We reconsidered the thermodynamic stability of Schwarzschild black hole taking mass-less quantum loops into account. At the one-loop level, we were able to analytically construct the partition function relying on various techniques. We utilized the non-local heat-kernel expansion, the underlying KS structure of Schwarzschild solution and lastly an appropriate expansion in curvatures. The results we presented are complete up to quadratic order in curvatures. Within our approach, it was quite straightforward to uncover the structure of the higher curvature operators that were ignored in our analysis.

The free energy shows a strong dependance on the particle content of the theory and two distinct cases emerge. If gauge fields dominate the theory, thermal stability is achieved at a mass proportional to $\sqrt{N_V} M_{\text{P}}$. The latter conclusion is insensitive to the short-distance details of quantum gravity. Otherwise, as in the standard model, stability is never achieved but the free energy develops a minimum at which the quantum-corrected entropy vanishes. Although it is not entirely clear what the latter result entails, it may certainly point to a stable ultra-Planckian remnant as a possible end state of Hawking evaporation\footnote{This possibility could be motivated by the generic properties of remnants, see \cite{Che2014} for a review.}. We also showed that our conclusions are largely insensitive to the missing higher curvature/loop corrections.  

There is quite a few open questions that one hopes to address in the future. First, it is straightforward to employ our formalism to study the thermal stability of Kerr black hole. It is crucial to investigate if quantum loops could play a similar role in this case as well. Second, studying charged black holes could be attempted with our formalism albeit the fact that gauge fields possess non-trivial background solution. Lastly, it is fascinating that the system's entropy vanishes at perhaps sub-Planckian temperatures. The implications of this feature warrant a thorough consideration.


\section{Acknowledgments}
This work is supported by the Science Technology and Facilities Council (STFC) under grant number ST/L000504/1.      


\end{document}